\documentstyle[psfig]{l-aa}

\psfigurepath{/gagax5/evol1/TEX/ARTICLES/AGB97.1/}

\newcommand {\mass}[1] {$\rm #1\,M_{\sun}$}

\newcommand {\chem}[2] {$\rm{}^{#2}\kern-0.8pt#1$}
\newcommand {\reac}[6] {$\rm\,{}^{#2}\kern-0.8pt{#1}\,({#3}\,,{#4})
  \,{}^{#6}\kern-0.8pt{#5}\,$}

\makeatletter
\def\cal{\fam\tw@}
\makeatother

\begin{document}

\title{The Brightest Carbon Stars}

\author{C.~A.~Frost \inst{1,2}, R.~C.~Cannon \inst{1,3},  J.~C.~Lattanzio
  \inst{1,2,4},  P.~R.~Wood \inst{5}, M.~Forestini \inst{4}}

\offprints{J. Lattanzio, Department of Mathematics, Monash University, 
  Clayton, Victoria, 3168, Australia}

\institute{
  Institute of Astronomy, Madingley Road, Cambridge, CB3 0HA, U.K. \and
  Department of Mathematics, Monash University, Clayton, Victoria,
     3168, Australia \and
  Department of Physiology and Pharmacology, School of Biological Sciences,
     University of Southampton, SO16 7PX, U.K. \and
  Laboratoire d'Astrophysique, Observatoire de Grenoble,
     Universit\'e Joseph Fourier, BP 53, F--38041 Grenoble Cedex 9, France
     \and
  Mount Stromlo and Siding Spring Observatories, The Australian National
  University, Private Bag, Weston Creek P.O., A.C.T. 2611, Australia
}

\date{Received ; Accepted }

\thesaurus{08.01.1;08.03.1;08.03.2;08.05.3;08.16.4}

\maketitle

\markboth{The Brightest Carbon Stars} {Frost et al.}

\begin{abstract}
  
  It is currently accepted that Hot-Bottom-Burning (HBB) in
  intermediate-mass asymptotic giant branch (AGB) stars prevents the
  formation of C~stars. Nevertheless, we present in this paper the results
  of some detailed evolutionary calculations which show that even with HBB
  we obtain C~stars at the highest luminosities reached on the AGB.
  This is due to mass-loss reducing the envelope mass so that HBB ceases
  but dredge-up continues. The high mass-loss rate produces an optically
  thick wind  before the star reaches C/O$>1$.
  This is consistent with the recent results of van Loon et al.
  (1997a,b) who find obscured C~stars in the Magellanic Clouds at
  luminosities up to $M_{bol} = -6.8$.


\end{abstract}

\section {AGB Evolution:~Mass-Loss and Dredge-Up}

AGB stars present a significant
challenge to theorists because they combine many physical
processes
which
are not well understood, such as mass-loss and
third dredge-up. Both are firmly established theoretically and
observationally, but reliable calculations are still impossible at
present (Vassiliadis \& Wood 1993, hereafter VW93; 
Frost \& Lattanzio 1996a).  Nevertheless,
we do have a qualitative understanding of these fascinating stars and the
main physical processes which govern their evolution (for recent reviews
see Frost \& Lattanzio 1996b, Lattanzio et al. 1996) as well as their
extensive nucleosynthesis (for example, see Sackmann \&
Boothroyd 1992, Gallino et al. 1996, Forestini \& Charbonnel 1997).

In the absence of Hot-Bottom Burning (HBB), that is for AGB stars initially
less massive than about 4 to \mass{5} (depending on metallicity $Z$), the
question of whether a star actually becomes a C~star or not depends
primarily on two things:

\begin{itemize}
  
\item the extent and time-variation of mass-loss on the AGB;
\item the efficiency (i.e. the depth) of dredge-up.

\end{itemize}

Mass-loss is usually included by fits
chosen to simulate observed rates.  For reliable models
we require a self-consistent scheme for determining the mass-loss rate and
its variation along the AGB and beyond.  There are
observational indications that toward the end of their evolution, some AGB
stars experience a short phase of extremely rapid mass-loss, often called a
``super-wind'' (Justtanont et al. 1996, Delfosse et al. 1997). These stars
are often surrounded by a dense circumstellar envelope and
are no longer visible in optical studies.  Some show at
present rather low mass-loss rates (Jura et al. 1988), suggesting a complex
time dependence.

The efficiency of dredge-up is usually described by the 
``dredge-up parameter'' $\lambda$
which is defined as the ratio of 
the mass dredged to the
stellar surface following a flash to the mass
processed by the hydrogen burning shell between two successive
shell flashes. There is no agreement concerning the value and dependence of
this parameter on time, stellar mass and composition.  It suffers
from many physical and numerical uncertainties (Frost \& Lattanzio 1996a),
and evolution calculations show that it varies greatly with luminosity,
mass and metal abundance (Wood 1996).
In synthetic evolution calculations (e.g.  Groenewegen \& de Jong
1993, Marigo, Bressan \& Chiosi 1996) it is usually assumed to be
a constant. 

Dredge-up and mass-loss are crucial because:

\begin{itemize}
  
\item the deeper the dredge-up the more \chem{C}{12} is added per pulse;
  
\item dredge-up alters the evolution of the star by 
  cooling the intershell region and changing the core-mass
  (e.g. VW93).

\item once dredge-up begins it occurs after each pulse, and the number of
  pulses depends critically on mass-loss which is the primary phenomenon
  determining the duration of the AGB phase (Sch\"onberner 1979);
  
\item mass-loss determines the envelope mass, and hence the amount of
  dilution which the dredged material experiences. The less dilution the
  sooner the star will become a C~star;
  
\item hydrostatic equilibrium of the envelope forces a given temperature at
  the bottom of the envelope, so that a minimum envelope mass is required for
  HBB to occur.

\end{itemize}

Synthetic AGB evolution calculations of Iben (1981) identified the
so-called ``Carbon star mystery'', namely that the then current models
predicted too many bright C~stars and not enough faint C~stars. The
importance of high mass-loss rates had been underestimated in these models,
and this is part of the explanation for the deficit of bright C~stars.
Recent models of low-mass stars by Straniero et al. (1995) have reduced, but
not eliminated, the discrepancy at low luminosity.

For AGB stars more massive than about 4 to \mass{5}, the occurrence of HBB
has been found to be the major factor affecting C~star formation (Sackmann
\& Boothroyd 1992).  In this phenomenon the bottom of the convective
envelope penetrates the top of the hydrogen burning shell so that some
nuclear processing occurs at the bottom of the envelope (during the
interpulse phase). This region is mixed throughout the photosphere and any
abundance changes produced by HBB can be directly observed at the surface.
This appears to provide a simple explanation for the high-luminosity Li-rich
AGB stars found by Smith \& Lambert (1989, 1990), as shown by Sackmann \&
Boothroyd (1992).  However it can also lead to CN cycling with the result
that the \chem{C}{12} added to the stellar envelope by dredge-up can then be
processed into \chem{C}{13} and \chem{N}{14}. If this CN cycle is operating
almost at equilibrium conditions, the star can avoid becoming a C~star
altogether. This was suggested some time ago by Wood, Bessell \& Fox (1983),
and was verified by Boothroyd, Sackmann \& Ahern (1993). Although this is
largely true, synthetic evolution models of Forestini \& Charbonnel (1997)
have recently suggested that a population of very bright C~stars can still
be produced even for stars experiencing strong HBB {\it if the mass-loss
  rate is not too high} (so that there are enough dredge-up episodes). It is
the aim of this Letter to confirm that very bright C~stars can indeed result
from a combination of effects which have revealed themselves when detailed
evolutionary models have been computed nearly all the way to the end of the
AGB evolution.

HBB requires a minimum envelope mass $M_e^{HBB}$ or temperatures will not
rise sufficiently at the base of the envelope. 
A minimum envelope mass $M_e^{TDU}$ is also required for third
dredge-up to occur, the exact value of which is currently unknown. 
If $M_e^{HBB} <M_e^{TDU}$ then the dredge-up will cease before 
HBB. If the opposite is
true, then dredge-up will continue after HBB has ceased. Note that
Groenewegen \& de Jong (1993) assumed that both processes stopped at the
same time (i.e. that $M_e^{HBB} = M_e^{TDU}$). From computations we have
performed, which will be reported fully elsewhere (Frost et al. 1997), it
appears that $M_e^{HBB} > M_e^{TDU}$, so that HBB ceases but dredge-up
continues.  This is, of course, relevant for the formation of high
luminosity C~stars.

\section {The brightest C~stars observed}

Recent observations by van Loon et al. (1997a,b) have been directed toward
finding stars with circumstellar envelopes in the Magellanic Clouds, in an
attempt to correct for the incompleteness of optical surveys at the highest
luminosities. They found 19 new objects and tried to determine which were
O-rich and which are C-rich.  Although this proved
impossible for some objects, they reached two important conclusions:

\begin{itemize}
  
\item the ratio $N_C/N_O$ of the number of C-rich to O-rich objects
  decreases with increasing luminosity. As the most massive AGB stars are
  also among the brightest ones, this is compatible with HBB effects;
  
\item this ratio {\it does not decrease to zero} even at the highest
  luminosities. This is inconsistent with a simple-minded understanding of
  HBB, but is consistent with our models, as we show below.

\end{itemize}

van Loon et al. (1997b) estimate that even at $M_{bol} = -7$, the value of
$N_C/N_O$ lies between 0.2 and 0.5\footnote{They suggest that the 
occurrence of both C-rich and O-rich stars at the same luminosity is 
possibly due to a mixture of compositions.  We believe it is more likely
due to a mixture of evolutionary stages.}.
Further, they find a C-rich object with
$M_{bol} = -6.8$, which makes it one of the most luminous AGB stars in the
Magellanic Clouds. Note that this is a C~star, but note also that the
brightest {\it optically visible\/} AGB 
stars in the Magellanic Clouds are in fact
O-rich (VW93). Thus the
conversion to a C~star appears to be related to the mass-loss which produces
the enshrouding. It is important for us to know the critical mass-loss rate
above which a star is no longer optically visible. As there are very few
visible stars with mass-loss above $10^{-6}$\mass{} /yr, and none above
$10^{-5}$\mass{} /yr 
(Gu\'elin, private
communication)  we shall use 
$5\times 10^{-6}$\mass{}/yr as the critical value.

\section {The brightest C~stars explained?}

In this section we show some preliminary results from a study of AGB
evolution of 4, 5 and \mass{6} models for Magellanic Cloud compositions
($Z=0.004$ and $0.008$) as well as solar ($Z=0.02$).  The models will be
discussed fully elsewhere (Frost et al. 1997). Calculations were
performed with the Mount Stromlo Stellar Evolution Code, with OPAL opacities
(Iglesias \& Rogers 1993) and mass-loss rates from VW93\footnote{We have not
included here the modification to the
formula for $M > $\mass{2.5} that delays the onset of the super-wind.}. 
Calculations of dredge-up have been performed 
as recommended by Frost \& Lattanzio (1996a) and include the entropy
adjustment of Wood (1981). We use a post-processing nucleosynthesis code to
follow the composition changes of 74 species up to sulphur within the
stellar models.  Each was evolved from before the ZAMS through many thermal
pulses, until rapid convergence problems occurred.

Here we discuss only the \mass{6} cases with compositions appropriate
to the Magellanic Clouds. We
computed 68 and 92 pulses for the $Z=0.008$ and $0.004$
cases, respectively.  Let \chem{C}{13}/\chem{C}{12} be
the number ratio $n($\chem{C}{13})$/n($\chem{C}{12}$)$ and C/O =
[$n$(\chem{C}{12})+$n$(\chem{C}{13})]/[$n$(\chem{O}{16})+$n$(\chem{O}{17})+
$n$(\chem{O}{18})], consistent with the observational definition
based on molecular studies.

\begin{figure}
  \psfig{file=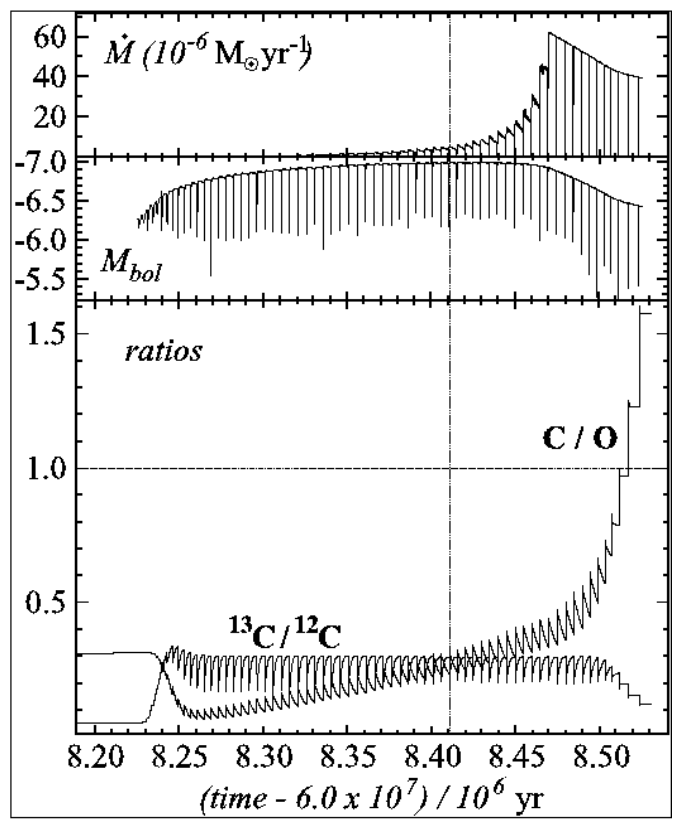,width=8.8cm}
  \caption[ ]{\label{M6Z008}
     AGB evolution of our \mass{6} model with $Z=0.008$. The top frame shows
     the mass-loss rate, the middle frame gives $M_{bol}$, and the lower
     frame presents the  \chem{C}{13}/\chem{C}{12} and C/O ratios. The
     time when the mass-loss rate exceeds our critical value
     of $5\times 10^{-6}$\mass{}/yr is shown as a vertical 
     line. We also show, by a horizontal
     line, the critical value C/O=1.}
\end{figure}

After each dredge-up event the \chem{C}{12} abundance is increased.  During
the following interpulse phase HBB transforms \chem{C}{12} into \chem{C}{13}
and \chem{N}{14}, driving
the \chem{C}{13}/\chem{C}{12} ratio toward its
equilibrium value of about 0.3 and decreasing the C/O value.
These effects are clearly seen in Fig. \ref{M6Z008} which shows the results for
the \mass{6} case with $Z=0.008$, appropriate to the Large Magellanic Cloud.
The C/O ratio rapidly drops from
the initial (pre-thermally pulsing) value of 0.31 as soon as the envelope
bottom temperature produces HBB and reaches a minimum value of about 0.07.
With each dredge-up event we see the increase in \chem{C}{12} produced by
strong dredge-up (we find $\lambda \sim 0.9$) but during the subsequent
interpulse phase this \chem{C}{12} is burned into \chem{C}{13} and
\chem{N}{14} by HBB, so that the \chem{C}{13}/\chem{C}{12} ratio remains at
its equilibrium value.  Note that as the evolution proceeds, the C/O ratio
begins to rise again.  Initially this is simply due to the large amount of
\chem{C}{12} added to the envelope. However during the later pulses the
decreasing envelope mass means less dilution of the dredged-up material
following each pulse as well as a decrease in the peak 
temperature at the bottom of
the convective envelope. In fact, we find that HBB ceases four or 
five pulses from
the end of the calculations. From that time, the C/O ratio climbs very
rapidly.
The model passes through C/O=1 and continues up to 1.5 at the
time that calculations ceased. We find $M_e^{HBB}$
exceeds $M_e^{TDU}$ and we determine  $M_e^{HBB}
\simeq $ \mass{1.98}. 
Except for the last few pulses, after HBB has ceased,
the \chem{C}{13}/\chem{C}{12}
ratio remains at its equilibrium value.  Note that this model 
reaches a peak of $M_{bol}= -7$ but that this has declined to $-6.6$ when
the star becomes a C~star.
Boothroyd, Sackmann \& Ahern estimate that HBB should
prevent the appearance of C~stars brighter than $M_{bol} = -6.4$,
which the current model shows is not always the case.

If a star is assumed to be no longer optically visible when its mass-loss
rate exceeds $5\times 10^{-6}$\mass{}/yr then our model would disappear
from optical surveys at an age of $6.841\times 10^7$ yr, as shown in
Figure~1,
when it is still clearly O-rich. This model only appears as a C-star
after it has become heavily enshrouded. This is consistent with the
observations of van~Loon et al (1997a,b), and is due to the fact that the
mass-loss rate determines three crucial things:~the termination of HBB, the
dilution of dredged-up material in the envelope, and the time at which the
star becomes no longer optically visible. The order of these events is
crucial.

\begin{figure}
  \psfig{file=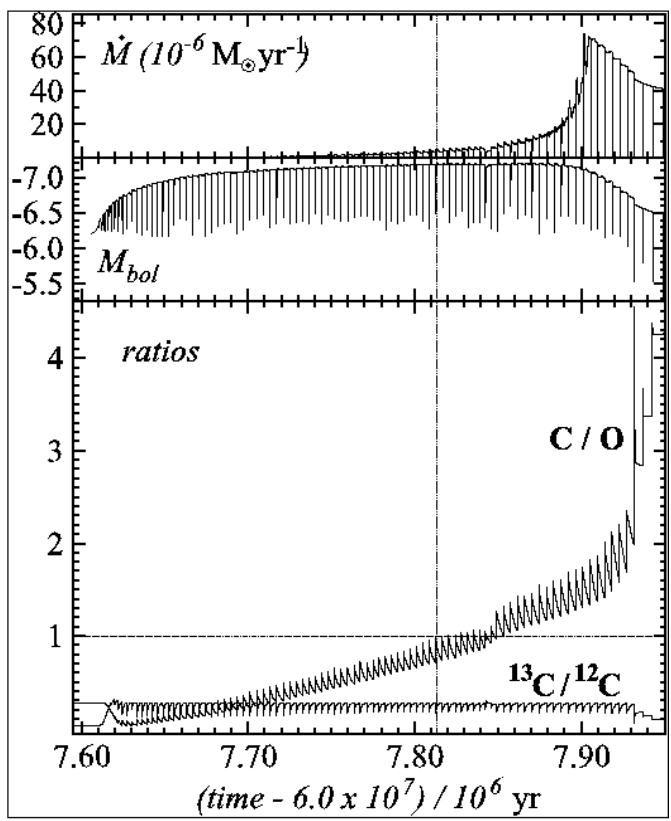,width=8.8cm}
  \caption[ ]{\label{M6Z004}
    Same as Fig. \ref{M6Z008} but for 
    $Z = 0.004$. Note that the two ``glitches'' in the abundances seen
    at $t=6.785$ and $6.793 \times 10^7$ yr are the result of
    ``degenerate pulses'', which will form the basis of a separate paper.}
\end{figure}

Figure \ref{M6Z004} shows the $Z=0.004$ case. Here again we see the continual
rise in the C/O ratio.
For this composition (appropriate to the Small
Magellanic Cloud) the C/O ratio exceeds unity long before HBB ceases. This
model is both a C~star and \chem{C}{13}-rich, which would see it classified
as a J~star.  When HBB ceases, the envelope mass is $M_e^{HBB} \simeq$
\mass{2.02}.  The luminosity when the model became a C~star 
was $M_{bol} =-7.2$, with post-flash dips decreasing this to $-6.3$.
Taking $5\times 10^{-6}$\mass{}/yr as the critical mass-loss rate for an
enshrouded star, we find that this model would no longer be visible for 
$t > 6.781\times 10^7$ yr, when C/O $\simeq 0.8$. 
The probability of observing a C~star is proportional to the fraction of the
interpulse period which the stars spends with C/O$>1$. Hence
this model is not a C~star while visible, but becomes a C~star 
soon after after it drops from
visibility. 

The two models presented show clearly that HBB prevents optically
visible C~stars from forming, but that mass-loss can then hide the star,
extinguish HBB, and permit it to become a (heavily obscured) C~star.
This is entirely consistent with the observations of AGB stars 
in the Magellanic Clouds.

\section {Discussion and prospects}

These full evolutionary models for AGB stars reveal a very complex interplay
between mass-loss, dredge-up and HBB. The carbon isotopic ratio is very
sensitive to these processes. Consequently, it is not currently possible to
confidently predict its evolution at the surface of AGB stars of various
initial masses and metallicities.

The intermediate-mass C~stars shown in this work
become C~stars earlier in their evolution as the metallicity is
decreased. Similarly, the time (and luminosity range) over which they show
high \chem{C}{13}/\chem{C}{12} also increases with decreasing $Z$.  We
finally note that, for the two cases shown here, HBB is terminated before
dredge-up ends.  This was also found to be the case for the solar
metallicity case (not reported here), with $M_e^{HBB} \simeq $\mass{2.57}.

These very bright C~stars, dredging-up material enriched in
\chem{C}{12} while undergoing strong HBB, substantially
increase their \chem{N}{14} envelope abundance. Compared to the beginning of
the AGB phase, the \chem{N}{14} enhancement factor in the wind of these
C~stars is about 4, 13 and 40 for the $Z = 0.02$, 0.008 and
0.004 cases respectively, which would make these intermediate-mass stars
significant producers of {\it primary\/} \chem{N}{14}.

The present evolution models of AGB stars also reveal that the
dredge-up depth together with both the rate and time-variation of the 
mass-loss are crucial
quantities deciding the final envelope composition of an AGB star (and its
planetary nebula). In particular, they determine whether such stars become
C~stars or not, and whether they are optically visible or not. 
The transformation of the
convective envelope (and wind) from O-rich to C-rich is made even more
complicated by HBB for the higher mass stars. Both mass 
loss and dredge-up depth are very 
uncertain, and the calculations reported here are extremely computationally
intensive. Only synthetic evolution, based on the results of
these detailed models, can investigate more fully the 
possible ranges of C/O
and \chem{C}{13}/\chem{C}{12} that can be found on the AGB.
This work is in progress.

Note that calculations with diffusive mixing (Herwig et al 1997)
produce different amounts of dredge-up and different
intershell conpositions to canonical calculations.
The implications of this are yet to be determined.

\section {Conclusion}
Regardless of the details of the
evolutionary calculations, we only expect to see luminous, massive
C~stars when there has been a significant reduction in the 
envelope mass, so that the HBB has stopped turning the 
\chem{C}{12} into \chem{N}{14} 
and the dredged-up \chem{C}{12} is not highly diluted in a large 
envelope.
Since the envelope mass only reduces
significantly when the super-wind is operating, we expect all the
massive, luminous C~stars to be dust-enshrouded.

\begin{acknowledgements}
  This work was supported by grants from the Australian Research Council,
  the British Council and the GDR ``Structure Interne des Etoiles et des
  Plan\`etes G\'eantes'' (CNRS). JCL would like to thank the
  Laboratoire d'Astrophysique (Grenoble) for its hospitality. JCL and MF
  also thank the Carthusian Monks for inspiration.
\end{acknowledgements}

\end{document}